\documentclass[pra, twocolumn, superscriptaddress, groupedaddress]{revtex4}
\pdfoutput=1

\usepackage[T1]{fontenc}
\usepackage{microtype}
\usepackage{amsmath}
\usepackage{amssymb}
\usepackage{graphicx}
\usepackage{subfloat}
\usepackage{cleveref}
\usepackage{natbib}
\usepackage{filecontents}

\usepackage{float}
\usepackage{color}

\newcommand{\abs}[1]{\left| #1 \right|}
\newcommand{\avg}[1]{\left< #1 \right>}
\renewcommand{\kappa}{\varkappa}

\DeclareMathOperator{\sech}{sech}

\begin{document}

\title{Dispersive radiation and regime switching of oscillating bound solitons in twin-core fibers near zero-dispersion wavelength}

\author{Ivan Oreshnikov}
\affiliation{ITMO University 197101, Kronverskiy pr. 49, St. Petersburg, Russian Federation}
\author{Rodislav Driben}
\affiliation{Department of Physics and CeOPP, University of Paderborn, Warburger St. 100, D-33098 Paderborn, Germany}
\author{Alexey Yulin}
\affiliation{ITMO University 197101, Kronverskiy pr. 49, St. Petersburg, Russian Federation}

\date{\today}

\begin{abstract}
We study resonant radiation generated by bound solitons in a twin-core fiber near zero-dispersion wavelength, in the presence of higher order dispersion terms. We propose a theoretical description of dispersive wave generation mechanism and derive resonance conditions. The presence of third order dispersion term leads to generation of polychromatic dispersive radiation and transition from the regime of center of mass oscillations to the regime of amplitude oscillations. Such a transition is not reproduced in the case of symmetric fourth order dispersion.
\end{abstract}

\maketitle

\section{Introduction}

\begin{figure*}[!th]
  \centering
  \includegraphics{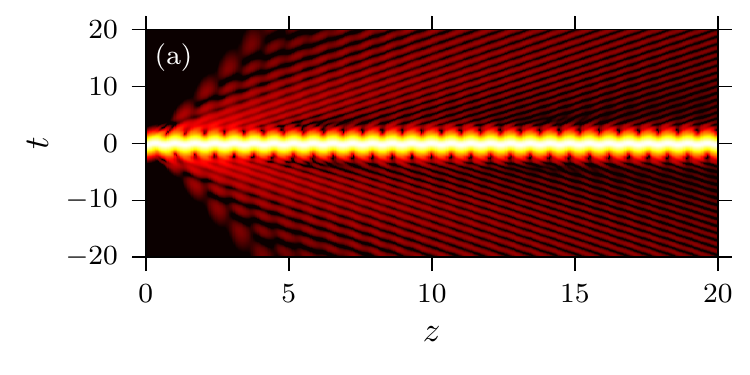}
  \includegraphics{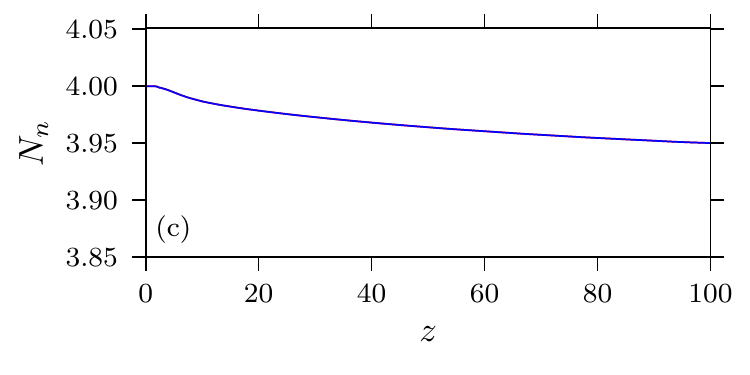}
  \includegraphics{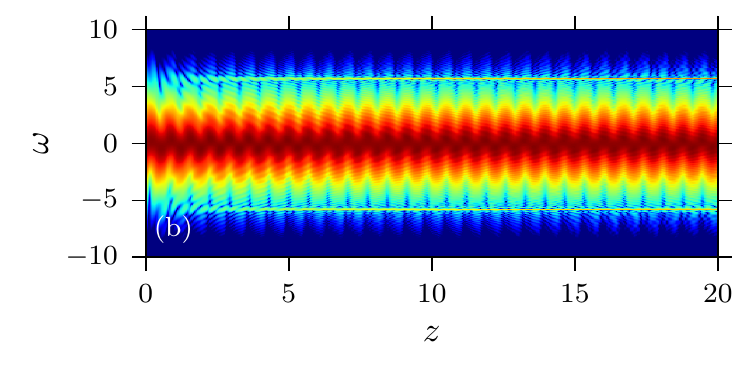}
  \includegraphics{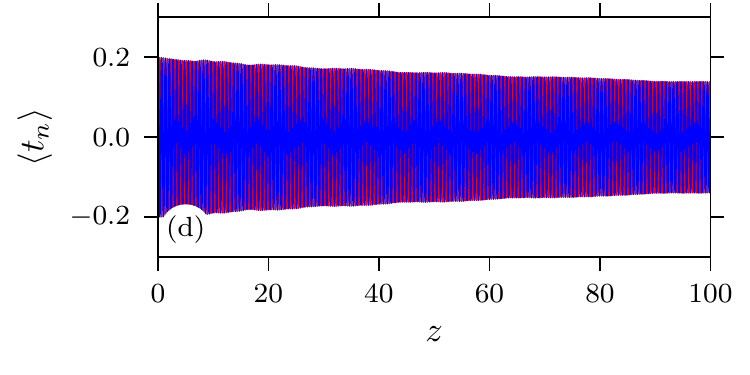}
  \caption{Second order dispersion case. (a) Time domain and (b) frequency domain plots of the field intensity in the first channel of the coupler $\left|u_{1}(z, t)\right|^{2}$ in logarithmic scale. The field in the second channel $\left|u_{2}(z, t)\right|^{2}$ is symmetric with respect to $t=0$. (c) and (d) are number of photons $N_n$ and central soliton positions $\avg{t_n}$ for both channels.}
  \label{fig:SODTimedomain}
\end{figure*}

Radiative effects of solitary waves propagating in nonlinear waveguides with high order dispersion has been attracting much of attention for long time \cite{KivsharMalomed_review}. This has two-fold motivation. On one hand this problem is interesting from the point of view of fundamentals of nonlinear waves dynamics. On the other hand this research has practical importance. Optical supercontinuum generation is one of the most prominent example of practical application of solitary waves \cite{GorbachSkryabin_review}.

In optics context one of the most interesting phenomenon is Cherenkov emission of dispersive wave by solitons propagating in nonlinear fibers with high order dispersion. This effect was reported back to 1986 \cite{Menyuk_Cherenk} and described in detail in \cite{akhmediev1995cherenkov}. A similar effect of transitional radiation of solitons moving in periodical systems was considered in \cite{Yulin_transitional}.

The interest to the interaction between the dispersive waves and optical solitons was revived when it occurred that it plays crucial role in optical supercontinuum generation  \cite{GorbachSkryabin_review}. In particular it was found that Cherenkov radiation affect the parameters of solitary waves and can even compensate for Raman self-frequency shift \cite{Skryabin_Science},\cite{Biancalana_AdiabaticTheory}. The combined effect of the soliton frequency shift and the resonant radiation modifies the spectrum of the output radiation strongly.

The interaction of solitons with the dispersive waves, either emitted by the solitons or the residual radiation of the initial pulse, leads to further enrichment of the spectrum of the output signal. The theory of this interaction was developed in \cite{Yulin_4wm_OL},\cite{Skryabin_4wm_PRE} and verified experimentally in \cite{Efimof_exp_1,Efimov_4wScattering,Efimov_PhaseSensScattering}. The role of the four-wave mixing of the solitons and dispersive waves in supercontinuum generation was revealed in a number of works \cite{Gorbach_NatureP, Babushkin, Yulin_trapping, Driben_trappingSC, driben2010supercontinuum, demircan2011controlling}.

Resonant radiation of oscillating solitary waves started with the analysis of the radiation of solitons propagating in active fibers with periodically varying linear gain \cite{Kodama_ResRad_ActivFiber}. Recently an analogous effect were reported for conservative systems with periodically varying parameters \cite{conforti2015parametric} and for oscillating solitons propagating in spatially uniform single-  \cite{driben2015resonant} and multi-mode \cite{wright2015ultrabroadband, krupa2016observation} fibers and in filament light bullets \cite{bree2017regularizing}.
	
The aim of this paper is to consider resonant radiation of bound state of solitons propagating in coupled fibers with high order dispersion. We show that the oscillations of the solitons results in the generation of polychromatic radiation with discrete spectrum and discuss how this radiation affects the dynamics of the bound state of the solitons. 

A system of nonlinear Schr\"odinger equations is a widely accepted model describing propagation of light in linearly coupled optical fibers with higher-order dispersion 
\begin{equation}
  \label{eq:EvolutionEquation}
  \begin{gathered}
  i \partial_{z} u_{n} +
    \hat D(i \partial_{t}) u_{n} +
    \abs{u_{n}}^{2} u_{n} +
    \kappa u_{m} = 0
  \\
  n, m = 1, 2 ~\text{and}~ n \ne m,
  \end{gathered}
\end{equation}
where $\kappa$ is a coupling parameter and $\hat D(i \partial_{t})$ is a dispersion operator defined by
\begin{equation*}
  \hat D(i \partial_{t}) =
      \sum \limits_{n=2}^{\infty}
      \frac{\beta_{n}}{n!} \left(
        i \partial_{t}
      \right)^{n}.
\end{equation*}

The equations (\ref{eq:EvolutionEquation}) have a rich plethora of linear and nonlinear solutions, In spite of the fact that the equations have been actively studied for many decdes, new interesting solutions have been being found even in the linear limit, see for example a recent work \cite{driben2014coupled} where optical Airy breathers were reported. 

For the purposes of this paper it is important that twin-core fibers support a family of symmetric and anti-symmetric solitons. However, as the stability analysis shows, both of those families are unstable after a certain threshold in $\eta / \kappa$ is reached (soliton amplitude to coupling constant ratio) and evolve into so-called asymmetric A- and B-states \cite{akhmediev1993novel, soto1993stability, akhmediev1994propagation}. 

An extensive research into adiabatic quasi-particle theory provides more insights into dynamics of solitons interaction. Launching the soliton into a single core of the fiber can, after a power threshold is reached, lead to a periodic switching \cite{chu1993analytical, chu1993soliton}. The symmetric solution with equal amplitude solitons of the same phase can be considered as an equilibrium point \cite{kivshar1989interaction}. Small perturbation of the solution either due to one of the solitons being delayed with respect to another, or uneven soliton amplitudes can lead to soliton parameters oscillations near that equilibrium. For example, a couple of delayed solitons can form a bound state with the solitons oscillating near a common center of mass \cite{kivshar1989interaction, abdullaev1989dynamics}. High-order dispersion and Raman effects alter the stability of the symmetric and antisymmetric solutions \cite{smyth1997dispersive,umarov1999soliton}, however, stable oscillating soliton states can be found in the presence of high order dispersion as well.

In the present paper we provide a theoretical description of resonant radiation generated by an oscillating pair of bound solitons. We also report a transition in the oscillation regime of coupled solitons from center-of-mass oscillations to periodic energy exchange between the channel, which is only apparent in the setting of an asymmetric dispersive characteristic of the medium.

\section{Resonance condition}

\begin{figure*}[!th]
  \centering
  \includegraphics{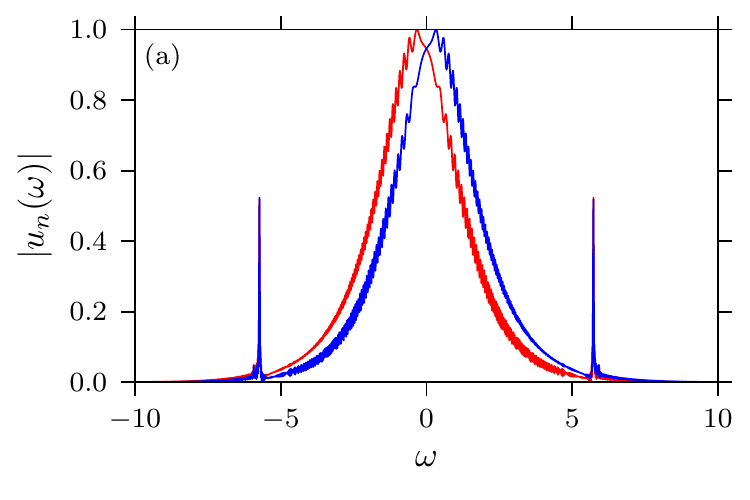}
  \includegraphics{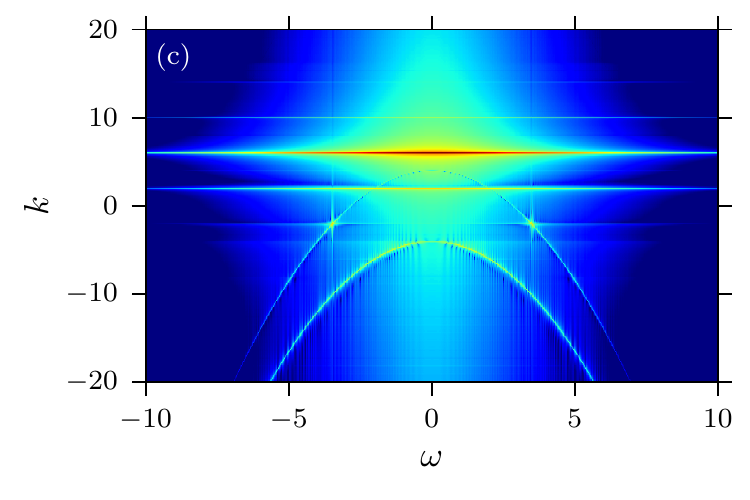}
  \includegraphics{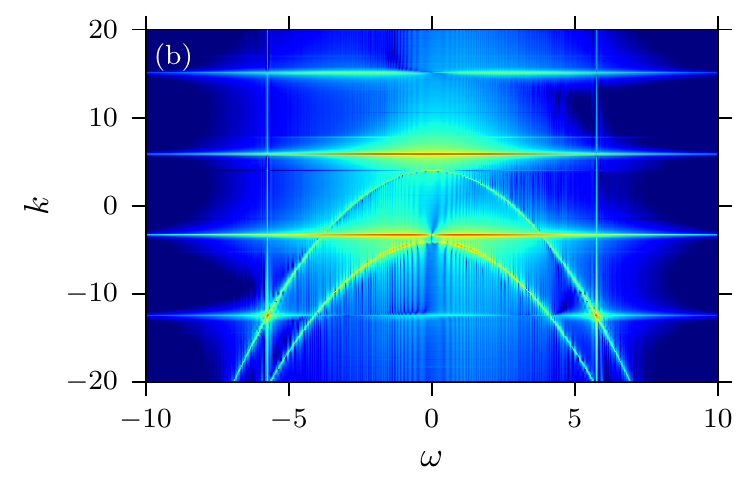}
  \includegraphics{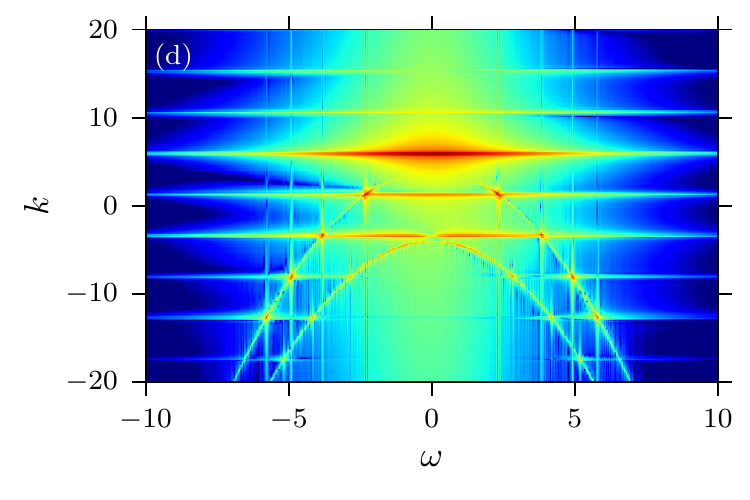}
  \caption{Second order dispersion case. (a) Output spectrum for a coupled state with initial delay of $2 \tau_0=0.4$ taken at $z=50$. (b) $\omega$-$k$ diagram of the field in the first channel $u_{1}(z, t)$ for $\eta_1 = \eta_2 = 2$; (c) --- the same diagram for the case of pure amplitude oscillations $\eta_1 = 2.1$ and $\eta_2 = 1.9$, $2 \tau_0 = 0$; (d) --- mixed regime at $\eta_1 = 2.1$, $\eta_2 = 1.9$, and $2 \tau_0 = 0.4$.}
  \label{fig:SODSpectrum}
\end{figure*}

We are interested in the evolution of two coupled solitons with their parameters periodically changing, either in the regime of amplitude oscillations, when the solitons in the neighbouring channels periodically exchange energy, or in the regime of center of mass oscillations, when both solitons sway around common center of mass. Both these configurations can arise from an initial condition
\begin{equation*}
  u_{1, 2}(z=0, t) = \eta_{1, 2} \sech\left( \eta_{1, 2} (t \pm \tau_0) \right), \\
\end{equation*}
where $\eta_{1, 2}$ are the amplitudes of the solitons and $2 \tau_0$ is an initial soliton delay.

An exact analytical solution to \cref{eq:EvolutionEquation} is not feasible. However, in the limit of weak coupling $\kappa \ll 1$ and in the absence of higher-order dispersion terms, an approximate quasi-particle solution can be constructed by use of a common \textit{ansatz} of a soliton with slowly changing amplitude $\eta_{n}(z)$, position $\tau_{n}(z)$, phase $\phi_{n}(z)$ and frequency $\omega(z)$
\begin{equation*}
  \begin{gathered}
  U_{n}(z, t) = A_{n}(z, t) \exp(i q z) \\
  A_{n}(z, t) =
    \eta_{n}
    \sech (\eta_{n} (t - \tau_{n}))
    \exp (- i \omega_{n} t + i \phi_{n}),
  \end{gathered}
\end{equation*}
where $q = \kappa + \eta^{2}/2$ is a wavenumber corresponding to a symmetric soliton solution. A concrete expressions for the slowly changing soliton parameters can be found either by means of perturbation theory \cite{abdullaev1989dynamics}, using integrals of motion \cite{kivshar1989interaction} or variational approach \cite{smyth1997dispersive}. As an example, let us consider symmetric center of mass oscillations, with both amplitudes being equal and constant $\eta_{1} = \eta_{2} = \eta$, and soliton positions $\tau_{n}(z)$ being periodic functions of $z$ with period $Z_{\tau}$. The solitons are oscillating out-of-phase, $\tau_{n}(z + Z_{\tau}/2) = - \tau_{n}(z)$, which also means that
\begin{equation}
  \label{eq:Swapping}
  A_{1}(z, t) = A_{2}(z + Z_{\tau}/2, t).
\end{equation}

To account for dispersive radiation we seek for the solution in the form of $u_{n} = U_{n} + \tilde u_{n}$, where $\tilde u_{n}$ is a small perturbation on top of an approximate quasi-particle solution $U_{n}$. Substituting this into \cref{eq:EvolutionEquation} and linearizing with respect to small $\tilde u_{n}$ we arrive at the equation
\begin{multline}
  \label{eq:PerturbationEquation}
  i \partial_{z} \tilde u_{n}
    + \hat D(i \partial_{t}) \tilde u_{n}
    + 2 \abs{U_{n}}^{2} \tilde u_{n}
    + U_{n}^{2} \tilde u_{n}^{*}
    + \kappa \tilde u_{m}
  \\
  = R[U_{n}, U_{m}] - \left(
    \hat D(i \partial_{t}) - 1/2 \, \partial^{2}_{tt}
  \right) U_{n}
\end{multline}
where $R[U_{n}, U_{m}]$ is a residue term left from substituting quasi-particle solution $U_{n}$ in the original equation \cref{eq:EvolutionEquation} in the absence of higher-order dispersion
\begin{equation*}
  R[U_{n}, U_{m}] =
    i \partial_{z} U_{n}
    + 1/2 \, \partial^{2}_{tt} U_{n}
    + \abs{U_{n}}^{2} U_{n}
    + \kappa U_{m}.
\end{equation*}

Far away from the solitons \cref{eq:PerturbationEquation} simplifies to $i \partial_{z} \tilde u_{n} + \hat D(i \partial_{t}) \tilde u_{n} + \kappa u_{m} = 0$ and admits to a plane-wave solution $u_{n} = a_{n} \exp\left(i k(\omega) z - i \omega t \right)$ with parameters
\begin{align}
  \label{eq:DispersiveRelations}
  k_{\pm}(\omega) = \hat D(\omega) \pm \kappa &&&
  [a_{1}, a_{2}]_{\pm} = [1,~ \pm 1].
\end{align}
This defines the asymptotics of the radiating states of the \cref{eq:PerturbationEquation}. The most important feature of the solution is that the upper ``$+$'' branch of the dispersive curve is symmetric, while the lower ``$-$'' branch is anti-symmetric \cite{arabi2016efficiency}. 

Thanks to the periodicity of quasi-particle solutions $U_{1, 2}(z, t)$ and property \eqref{eq:Swapping} we can represent it as a sum of spatial harmonics in the form of the following Fourier series
\begin{equation*}
  \begin{bmatrix}
    U_{1}(z, t) \\ U_{2}(z, t)
  \end{bmatrix}
  = \sum_{n}
  \begin{bmatrix}
    +1 \\ (-1)^{n}
  \end{bmatrix}
  C_{n}(t) \exp(i (q + n k_{0}) z),
\end{equation*}
where $k_{0} = 2 \pi / Z_{\tau}$ and $C_{n}(t)$ is defined by an integral $C_{n}(t) = (1/Z_{\tau}) \cdot \int_{-Z_{\tau}/2}^{+Z_{\tau}/2} A_{n}(z, t) \exp(- i n k_{0} z) \, dz$. Here we can note an important property in $[U_{1}, U_{2}]$, namely, that even spatial harmonics are symmetric and odd harmonics are antisymmetric. Similar representations can be written for both terms in the right hand's side of \cref{eq:PerturbationEquation}, $R[U_{n}, U_{m}]$ and $(\hat D - 1/2 \, \partial_{tt}^{2}) U_{n}$.

Dispersive radiation is generated due to a resonance between one of the source terms in the right hand's side and a radiating solution of \cref{eq:PerturbationEquation}, which happens when a wavenumber of a source term matches with a wavenumber of a radiating solution (see \cite{akhmediev1995cherenkov} for details). Due to the opposite symmetries of upper and lower branches of the dispersive curves, as well as even and odd harmonics of the source an additional restriction applies: the upper branch is excited by even harmonics and odd harmonics excite the lower branch. Following these statements we write the resonance conditions:
\begin{subequations}
  \label{eq:ResonanceConditions}
  \begin{align}
  \hat D(\omega) + \kappa &= 2n \cdot k_{0}       + q \\
  \hat D(\omega) - \kappa &= (2n + 1) \cdot k_{0} + q.
  \end{align}
\end{subequations}
In the case of a moving source $U_{1, 2}$ we can employ Galilean invariance of \cref{eq:EvolutionEquation} and change to solitons' reference frame. The resonance conditions in this case change to
\begin{subequations}
  \label{eq:ResonanceConditionsMovingFrame}
  \begin{align}
  \hat D(\omega) + \kappa &= 2n \cdot k_{0}       + q + v \omega \\
  \hat D(\omega) - \kappa &= (2n + 1) \cdot k_{0} + q + v \omega,
  \end{align}
\end{subequations}
where $v$ is the central velocity of the coupled solitons. This approach to resonance conditions is analogous to one used earlier for higher-order solitons \cite{driben2015resonant} and solitons in dispersion oscillating fibers \cite{conforti2015parametric}, but the situation is complicated by double dispersion curve and alternating symmetries in soliton harmonics and dispersive solutions.

\begin{figure*}[h]
  \centering
  \includegraphics{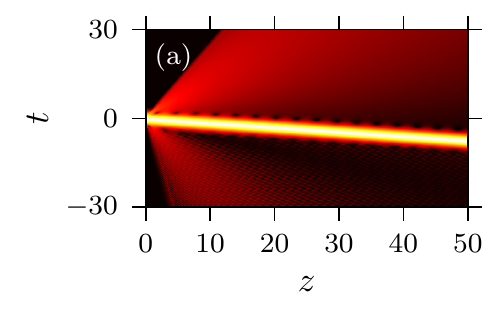}
  \includegraphics{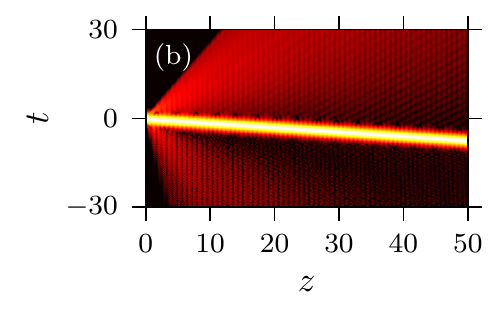}
  \includegraphics{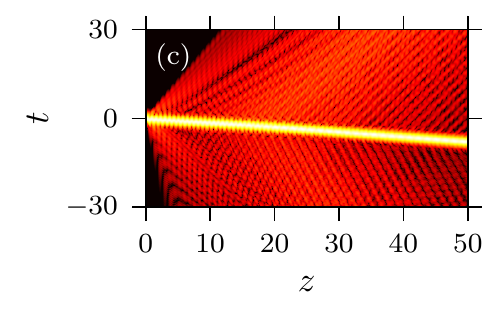}
  \caption{Third order dispersion case. Time domain plots of intensity distribution in the first channel $\abs{u_{1}}$ for the oscillating soliton states (with $\eta_1 = 2.1$ and $\eta_2 = 1.9$) at different initial values of relative delay $2 \tau_0$. (a) is the symmetric case $2 \tau_0 = 0.0$, (b) is the case of $2 \tau_0 = 0.1$, and (c) is the case of $2 \tau_0 = 0.4$.}
  \label{fig:TODTimedomain}
\end{figure*}

\begin{figure*}[h]
  \centering
  \includegraphics{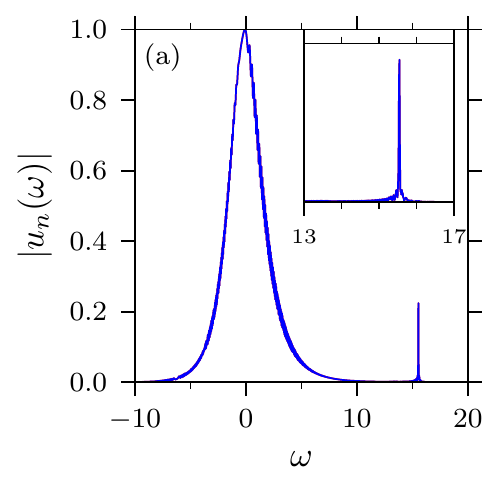}
  \includegraphics{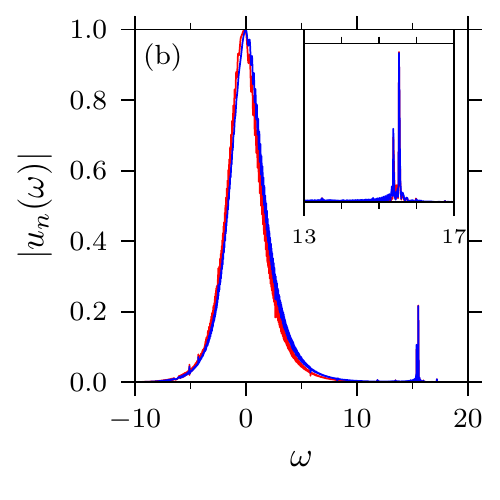}
  \includegraphics{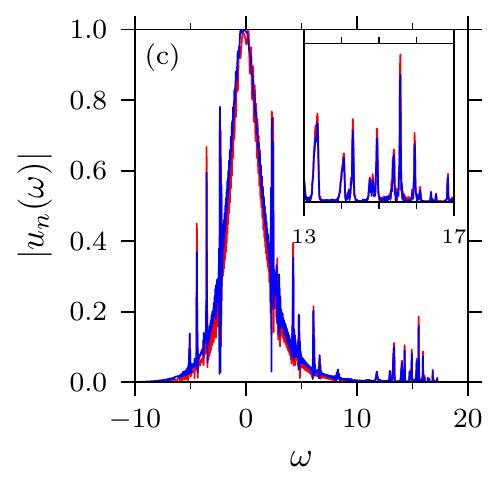}
  \includegraphics{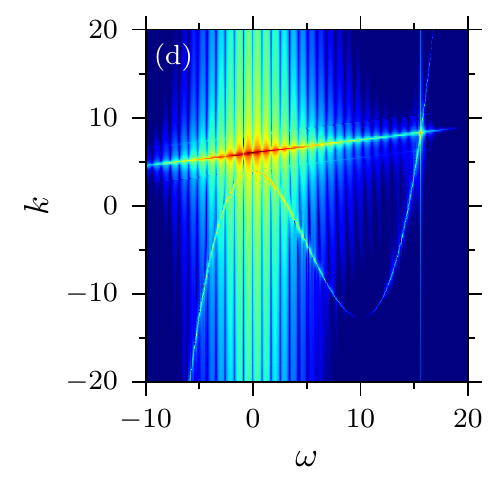}
  \includegraphics{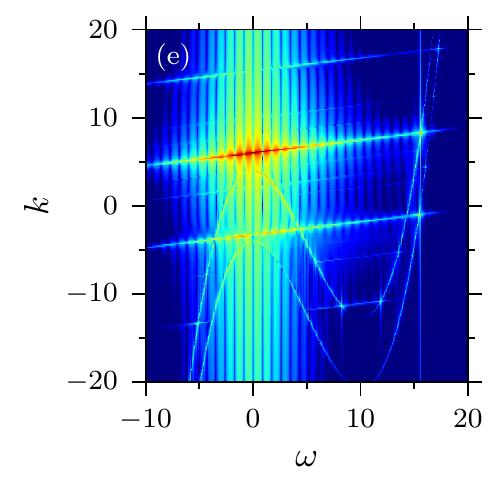}
  \includegraphics{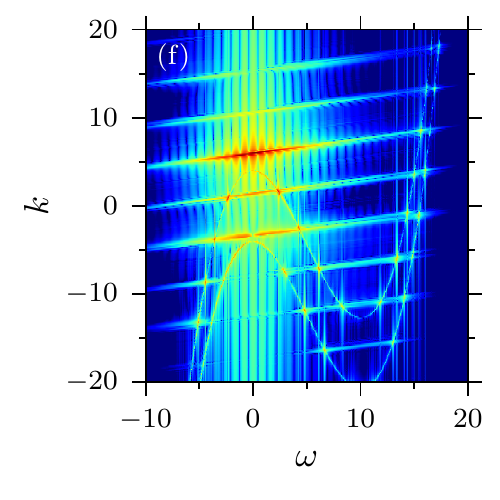}
  \caption{Third order dispersion case. Output spectra at $z = 0$ and $\omega$-$k$ diagrams for the oscillating soliton state (with $\eta_1 = 2.1$ and $\eta_2 = 1.9$) at different initial values of relative delay $2 \tau_0$. (a, d) is a symmetric case $2 \tau_0 = 0$, (b, e) is the case of $2 \tau_0 = 0.1$, and (c, f) is the case of $2 \tau_0 = 0.4$. Insets in (a, b, c) show the vicinity of Cherenkov resonance.}
  \label{fig:TODSpectrum}
\end{figure*}

\begin{figure*}[h]
  \centering
  \includegraphics{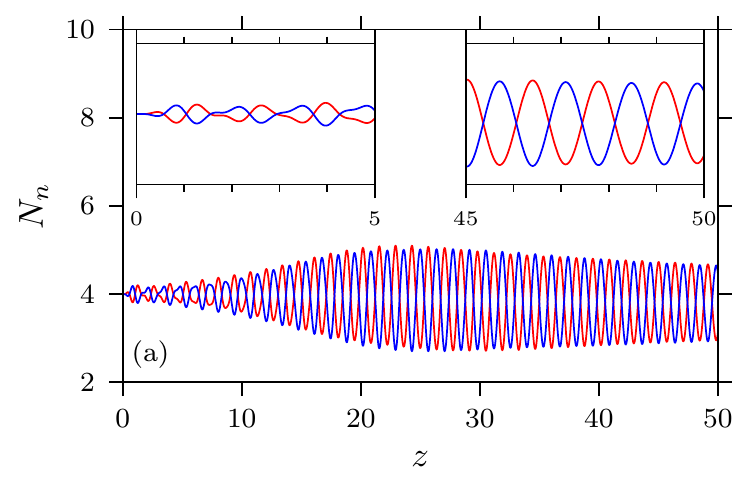}
  \includegraphics{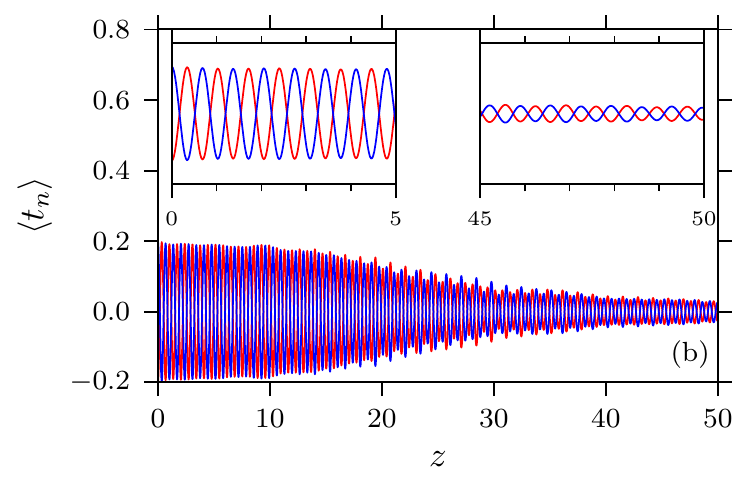}
  \caption{Third order dispersion case. (a) is the number of photons $N_{n}$ and (b) is the soliton position $\avg{t_{n}}$ for the initial conditions of equal amplitude solitons $\eta_1 = \eta_2 = 2$ with significant temporal delay $2 \tau_0 = 0.4$ (corresponds to Figures \ref{fig:TODSpectrum}c and \ref{fig:TODSpectrum}f).}
  \label{fig:TODMoments}
\end{figure*}

\clearpage

The reasoning above can be repeated for pure amplitude oscillations involving two solitons periodically exchanging energies between the channels without any temporal delay --- a state that evolves from the initial condition with $\eta_1 \ne \eta_2$ and $\tau_0 = 0$. Intermediate regime, featuring both oscillation in the soliton amplitudes $\eta_{n}$ and central position $\tau_{n}$ is not so tractable. We can suppose that in the intermediate regime the symmetry-based resonance exclusion would not hold, and all the resonances would contribute to the dispersive radiation.

\section{Second order dispersion}

We start our numeric analysis with a simple case of second order dispersion with $\hat D(i \partial_{t}) = 1/2 \, \partial_{tt}^{2}$. To make the radiation more prominent at shorter simulation distances we choose $\eta=2$. It is known from the bifurcation analysis \cite{akhmediev1993novel} that for $q \gtrapprox \sqrt{1.2 \kappa}$ the symmetric solution (in our case $2 \tau_{0} = 0$) is unstable and if launched evolves toward stable asymmetric A-type state. In order to avoid that we set the coupling $\kappa = 4$.

Launching the solitons with an initial relative delay $2 \tau = 0.4$ we observe propagation of a bound states with the solitons oscillating near the common center of mass (for a time domain plot of the first channel see \Cref{fig:SODTimedomain}a). We also see that the pair of solitons immediately begins to generate resonance radiation (both time and frequency domain plots in Figures \ref{fig:SODTimedomain}a and \ref{fig:SODTimedomain}b). Figures \ref{fig:SODTimedomain}c and \ref{fig:SODTimedomain}d display the first two moments of intensity distribution functions $\abs{u_{n}(z, t)}^{2}$, namely
\begin{subequations}
  \begin{align}
  N_{n}(z) &=
    \int \limits_{-T/2}^{+T/2}
      \abs{u_{n}(z, t)}^{2} \, dt \\
  \avg{t_{n}}(z) &=
    \frac{1}{N_{n}} \int \limits_{-T/2}^{+T/2}
      t \abs{u_{n}(z, t)^{2}} \, dt
  \end{align}
\end{subequations}
as functions of distance $z$, where the integrals are taken over a finite time window of $T = 20$. The integral $N_{n}$ is a characteristic of total energy attributed to a pulse, sometimes also called a \textit{number of photons}, and $\avg{t_{n}}$ is the central position of a pulse. In Figures \ref{fig:SODSpectrum}c and \ref{fig:SODSpectrum}d we can notice that due to shedding of dispersive radiation both solitons loose energy $N_{n}$. However, an additional decay in the amplitude of position $\avg{t_n}$ oscillations is present. This indicates that the internal oscillation mode loses energy due to radiation damping.

In the output spectrum the dispersive radiation manifests itself as two sharp peaks near $\omega \approx \pm 6$ (see \Cref{fig:SODSpectrum}a for the spectral density of both channels at $z=50$). To compare the simulated spectrum with the predictions of resonance conditions \eqref{eq:ResonanceConditions} we take the field $u_{1}(z, t)$ in the simulation domain and perform a 2-dimensional Fourier transform, moving into $\omega$-$k$ plane. The intensity plot of the resulting spectrum (\cref{fig:SODSpectrum}b) clearly shows two branches of dispersive curve (parabolas starting at $\pm \kappa$) as well as a set of soliton's spatio-temporal harmonics (horizontal line). The line passing through $k = 6$ is the fundamental $n = 0$ harmonics of the soliton ($k = q$). The horizontal line below it is the $n = -1$ harmonics of the soliton. It does intersect with the upper branch of the dispersive curve near $\omega \approx \pm 4$. However, due to the different symmetries of the solution these resonances do not contribute to the radiation. The horizontal line corresponding to $n = -2$ harmonics (faint, passing through $k = -13$) is in resonance with both the upper and the lower branches of the dispersive curve. Due to the difference in the symmetry, the resonance with the lower branch does not lead to generation of dispersive waves. The only contributing resonance is between the $n = -2$ harmonics of the soliton and the upper branch of the dispersive curve, which is indicated by the cross-hair pattern centered at the intersection points.

To confirm our reasoning about the soliton's harmonic and dispersive curves' parity we additionaly look at the case of pure amplitude oscillations with $\eta_1 = 2.1$, $\eta_2 = 1.9$ and no delay between the solitons $2 \tau_0 = 0$ (see \Cref{fig:SODSpectrum}c) and then the case of a maixed regime with $\eta_1 = 2.1$, $\eta_2 = 1.9$ and $2 \tau_0 = 0.4$ (see \Cref{fig:SODSpectrum}d). In the regime of pure amplitude oscillations the same parity selection rule holds, as is evident from no resonance between the $n=-1$ harmonic of the soliton and the upper branch of the dispersive curve. In the case of mixed oscillation regime ($\eta_{1} = 2.1$, $\eta_{2} = 1.9$ and $2 \tau_{0} = 0.4$) no harmonic is either purely symmetric or antisymmetric, rather all of them are asymmetric and thus every intersection of the soliton's harmonics with the dispersive curves contribute to the dispersive radiation.

\section{Third and fourth order dispersion}

\begin{figure*}[!th]
  \centering
  \begin{minipage}[b]{0.49\textwidth}
    \includegraphics{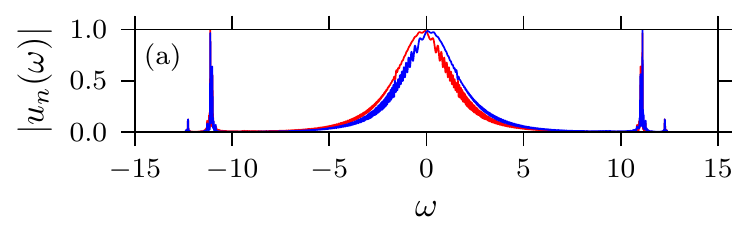}
    \includegraphics{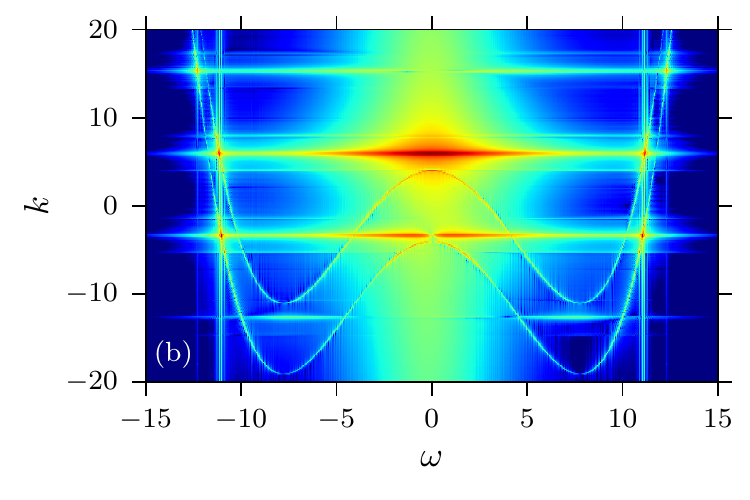}
  \end{minipage}
  \begin{minipage}[b]{0.49\textwidth}
    \includegraphics{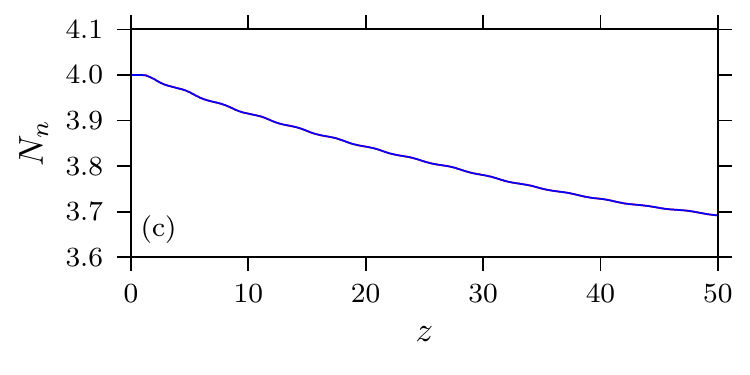}
    \includegraphics{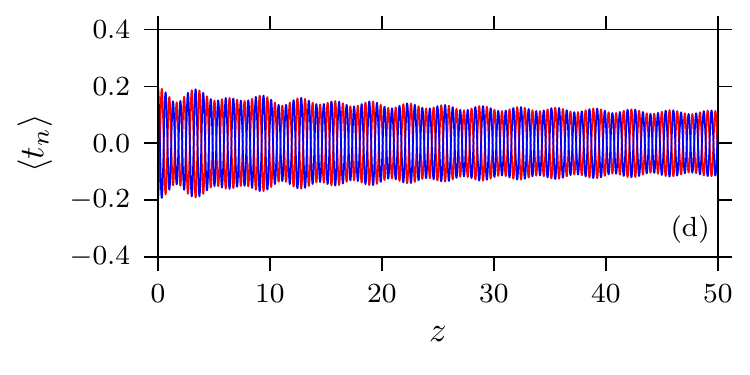}
  \end{minipage}
  \caption{Fourth order dispersion case. (a) Output spectrum for a coupled state with initial delay of $2 \tau_0=0.4$ taken at $z=50$. (b) $\omega$-$k$ diagram of the field in the first channel $u_{1}(z, t)$.}
  \label{fig:FOD}
\end{figure*}

We proceed with the case of third order dispersion using the operator $\hat D = -1/2 \, \partial_{t}^{2} - i \beta_{3}/6 \, \partial_{ttt}^{3}$ with $\beta_{3} = 0.2$. We probe three different relative delays $2 \tau_{0} = 0.0$, $0.1$, and $0.4$. The case of no relative delay $2 \tau_0 = 0$ behaves exactly as a single soliton of a scalar equation in the presence of third order dispersion perturbation, i.e. the soliton generates monochromatic resonant radiation at the frequency predicted by a resonance condition  as it is shown in Figures \ref{fig:TODSpectrum}a and \ref{fig:TODSpectrum}d. Introduction of a small relative delay $2 \tau_0 = 0.1$ leads to a splitting of a single spectral line into a tight frequency pair (Figure \ref{fig:TODSpectrum}b). This additional spectral line is due to contribution of the resonance between $n = -1$ harmonic of a soliton pair and the lower dispersive curve (compare with Figure \ref{fig:TODSpectrum}d). However, a larger delay (in our simulations $2 \tau_0 \ge 0.2$) reveals a more complicated dynamics. As an example, let us look at the output spectrum in Figure \ref{fig:TODSpectrum}c for the extreme case of $2 \tau_0 = 0.4$. Aside from the pronounced spectral lines near the spectrum spectrum of the solitons we can see that the vicinity of Cherenkov resonance is filled with a number of separated frequency pairs (inset in Figure \ref{fig:TODSpectrum}c). $\omega$-$k$ diagram in Figure \ref{fig:TODSpectrum}f indicates, that in addition to a set of harmonics corresponding to the center of mass oscillations of the soliton pair, the wavenumber spectrum contains a series of additional lines between the original harmonics. In addition to that, the resonance exclusion based on the symmetries does not work anymore, and every intersection between a dispersive curve and a soliton harmonic contributes to the dispersive radiation.

To track the origin of the additional spatial harmonics in the solitons' spectrum in Figure \ref{fig:TODSpectrum}d we look at the moments $N_{n}$ and $\avg{\tau_n}$ in Figure \ref{fig:TODMoments}. Insets in figures demonstrate input and output parts of the simulation. As it is evident from the plots, the center of mass oscillations (with period $Z_{\tau} \approx 0.7)$ decay rapidly (panel b). The system, however, does not evolve towards a steady state, but instead develops amplitude oscillations with period $Z_{\eta} \approx 1.4$ (panel a). There are known approximations for oscillation periods in both regimes provided that amplitudes are small \cite{kivshar1989interaction, abdullaev1989dynamics}. Unfortunately, neither weak coupling nor small amplitude approximations are applicable to our case.

To compare this to the case of symmetric fourth order dispersion we consider the dispersion operator $\hat D = -1/2 \, \partial^{2}_{tt} + \beta_{4}/24 \, \partial^{4}_{tttt}$ with $\beta_{4} = 0.1$. Launching the solitons with initial relative delay of $2 \tau_{0} = 0.4$ does not lead to the oscillation regime switching, see Figures \ref{fig:FOD}c and \ref{fig:FOD}d. And indeed the output spectrum and the $\omega$-$k$ diagram at \cref{fig:FOD}b is similar to the case of second order dispersion --- since the soliton . We can propose a preliminary hypothesis that in the presence of asymmetric dispersion profile center of mass oscillations of a soliton pair become unstable, while the amplitude oscillations are not affected by such a perturbation. The detailed analysis of the regime stability lies outside the scope of the current paper.

\section{Conclusions}

In this paper we have considered resonant radiation of oscillating bound states of solitons propagating in the coupled fibers with high order dispersion. The resonant condition is derived analytically and it is shown that the parity of the soliton bound state defines the parity of the radiated dispersive waves. An excellent agreement between the frequencies of the radiation predicted by the resonance condition and the positions of the spectral lines observed in direct numerical simulations is demonstrated.     

It is shown that the resonant radiation can be caused either by the oscillations of the mutual delay between the solitons or by periodic energy exchange between the solitons. The oscillations of the mutual delay and the soliton amplitudes have different periods and, consequently, lead to emission of the resonant radiation of different frequencies. The mutual delay and the amplitude oscillations can occur simultaneously resulting in the formation of reach radiation spectrum. By numerical simulations it was shown that the energy of the oscillations goes into the resonant radiation and finally a non-oscillating bound state of the solitons forms. The energy of the solitons in the bound state decreases if the condition of Cherenkov synchronism is fulfilled.  

It was observed and investigated numerically that the recoil from the resonant radiation can result in the drastic change of the dynamics of the bound state of the solitons. In particular in the case of symmetric (in the sense $k \rightarrow -k$) dispersion two identical solitons launched with a small delay in the first and in the second fiber exhibit oscillations of the soliton mutual delay but the amplitudes of the solitons are not changing during propagation. As it is mentioned above the oscillations slowly decay because of the radiation of the resonant mode. 

However in the case of asymmetric dispersion the oscillation of the soliton mutual delay decay much quicker and give rise to the quasi-periodical energy transfer between the solitons. In their turn the oscillations of the soliton amplitudes result in the radiation of resonant modes with different frequencies and relatively slow decay with the propagation distance.      

The reported results explains the dynamics of the bound states of the solitons in coupled nonlinear fibers with high-order dispersion and are important from the fundamental point of view and can possibly be used for better understanding of the process of supercontinuum generation in complex waveguiding systems, in particular in dual core silica fibers \cite{buczynski2009ultra}.

\begin{acknowledgments}
AY acknowledges support from the Government of the Russian Federation (Grant 074-U01) through the ITMO University early career fellowship.
\end{acknowledgments}

\bibliography{Bibliography}

\end{document}